\documentclass[10pt,letterpaper]{article}
\usepackage{amsmath}
\usepackage{opex3}
\usepackage{graphicx}
\usepackage{cite}
\usepackage{setspace}

\begin{document}


\title{Entanglement-based quantum key distribution with biased basis choice via free space}

\author{Yuan Cao,$^1$ Hao Liang,$^1$ Juan Yin,$^1$ Hai-Lin Yong,$^1$ Fei Zhou,$^1$ Yu-Ping Wu,$^1$ Ji-Gang Ren,$^1$ Yu-Huai Li,$^1$ Ge-Sheng Pan,$^1$ Tao Yang,$^1$ Xiongfeng Ma,$^{2,3}$ Cheng-Zhi Peng,$^{1,*}$ and Jian-Wei Pan$^1$}
\address{$^1$Shanghai Branch, National Laboratory for Physical Sciences at Microscale and Department of Modern Physics, University of Science and Technology of China, Shanghai 201315, China.\\
$^2$Center for Quantum Information, Institute for Interdisciplinary Information Sciences, Tsinghua University, Beijing 100084, China\\
$^3$xma@tsinghua.edu.cn\\
$^*$pcz@ustc.edu.cn\\
}

%
%



\begin{abstract}
We report a free-space entanglement-based quantum key distribution experiment, implementing the biased basis protocol between two sites which are $15.3$ km apart. Photon pairs from a polarization-entangled source are distributed through two 7.8-km free-space optical links. An optimal bias 20:80 between the $X$ and $Z$ basis is used. A post-processing scheme with finite-key analysis is applied to extract the final secure key. After three-hour continuous operation at night, a 4293-bit secure key is obtained, with a final key rate of 0.124 bit per raw key bit which increases the final key rate by $14.8\%$ comparing to the standard BB84 case. Our results experimentally demonstrate that the efficient BB84 protocol, which increases key generation efficiency by biasing Alice and Bob's basis choices, is potentially useful for the ground-satellite quantum communication.
\end{abstract}


\bibliographystyle{osajnl}

\section{Introduction}

Quantum key distribution (QKD) offers an approach of information-theoretically secure key extension between two remote parties, Alice and Bob, based on the fundamental principles of quantum mechanics. There are two types of QKD schemes: one is the prepare-and-measure scheme, such as the Bennett-Brassard-1984 (BB84) protocol \cite{Bennett:BB84:1984} and the Bennett-1992 protocol \cite{Bennett:B92:PhysRevLett}; the other is the entanglement-based scheme, such as the Ekert-1991 protocol \cite{Ekert:QKD:1991} and the Bennett-Brassard-Mermin-1992 (BBM92) protocol \cite{Bennett:BBM92:1992}. Since the first QKD experimental demonstration in the early 1990s \cite{Bennett:QKDExp:1992}, QKD has rapidly developed towards the stage for real-life applications \cite{commercialQKD}.

Generally speaking, there are mainly two types of quantum channels, fiber links and free-space links. Considering absorptive channel losses and detector dark counts, the distances of current fiber-based QKD experiments are limited to hundreds of kilometers \cite{DixonYuan:OE:2008,YuanNJP:ghzQKD:2009,Liu:10}. For free-space links, on the other hand, the photon losses and photon decoherence are almost negligible in outer space, thus the ground-satellite QKD offers a promising way to run large-scale quantum communication schemes \cite{IEEE:satellite:2003}. Along this line, the free-space quantum communication systems have been successfully demonstrated over large scale distances \cite{Zeilinger:144Decoy-state:2007,naturephys:144km:2007,PhysRevLett.94.150501}, with more practical and compact systems \cite{Erven:08}. Recently, a series of experimental demonstrations are presented for feasibility tests of ground-satellite quantum key distribution \cite{Nauerth:airtoground:2012,yangbin:decoystate:2013} and teleportation \cite{Yin:tele100km:2012,Maxiaosong:100kmtele:2012}.

In the BB84 protocol, Alice prepares qubits in the $X$ or $Z$ basis randomly and sends them to Bob, who measures randomly in one of the two bases. In the original proposal, Alice and Bob both choose two bases with equal probabilities (50:50). Thus, in the basis-sift procedure, they have to discard half of the detection events (i.e., the basis-sift factor is 1/2). Later, Lo, Chau and Ardehali proposed an efficient version of the BB84 protocol \cite{Lo:EffBB84:2005}, in which Alice and Bob choose two bases, the $X$ and the $Z$, unevenly (with a bias) to increase the basis-sift factor. In the infinite-key-size limit, this efficient BB84 protocol can improve the basis-sift factor from 1/2 to 1, compared to the original BB84. The efficient BB84 protocol can be naturally extended to its entanglement-based version, the BBM92 protocol \cite{Bennett:BBM92:1992}.

The first local experimental demonstration of the efficient BB84 in laboratory was presented in 2009 \cite{Erven:Biased:2009}, where the bias was simulated by introducing extra losses in one of the bases. In practice, such kind of biased setup cannot increase the QKD efficiency. In our experiment, the efficient BB84 protocol is demonstrated by actively changing the ratio of two bases, which can increase the final key rate by $14.8\%$ comparing to the standard BB84 case.

In this paper, we report a free-space entanglement-based QKD experiment using a parametric down-conversion (PDC) source. The distances between the PDC source and two receivers are both above $7.8$ km which is comparable to the effective thickness of the atmosphere. Furthermore, the efficient BB84 protocol \cite{Lo:EffBB84:2005} is implemented in our QKD experiment. In the evaluation of the final secure key, we apply a post-processing scheme with finite-key analysis \cite{Ma2011172,Finite:Long:10}.

In Section \ref{Sec:Biased:Setup}, we describe the experiment setup and implementation details. In Section \ref{Sec:Biased:Post}, we present the data postprocessing with finite-key analysis. We finally conclude in Section \ref{Sec:Biased:Conclusion} with discussions.

\section{Experimental implementation} \label{Sec:Biased:Setup}
We establish two free-space optical links among three sites, two receiver stations on two sides and one transmitter in the middle, near Qinghai Lake in China. The main purpose of this experiment is to implement a free-space entanglement-based QKD with the efficient BB84 scheme.

\subsection{Experimental setup}
At the transmitter, sited in the lawn of a local family near Qinghai Lake, the core equipment is a compact entanglement source ($45\times30~cm^{2}$) which utilizes a periodically poled $KTiOPO_{4}$ (ppKTP) crystal in a polarization-based Sagnac interferometer. By a continuous wave (CW) pump laser with a power of 12 mW, the PDC source can generate around $10^{7}$ entangled pairs per second with a central wavelength of 811 nm. In a 3 ns coincidence time window, about $5\times10^{5}$ photon pairs per second is detected locally with long-pass filters (with cutoff wavelengths at 488 nm). The full width at half maximum (FWHM) of the generated photon pairs is about 0.2 nm. Two commercial Kepler telescopes with 80 mm diameter was used to transmit the entangled photon pairs to two receiver sites, respectively.

\begin{figure}[hbt]
\centering
\includegraphics[width=13cm]{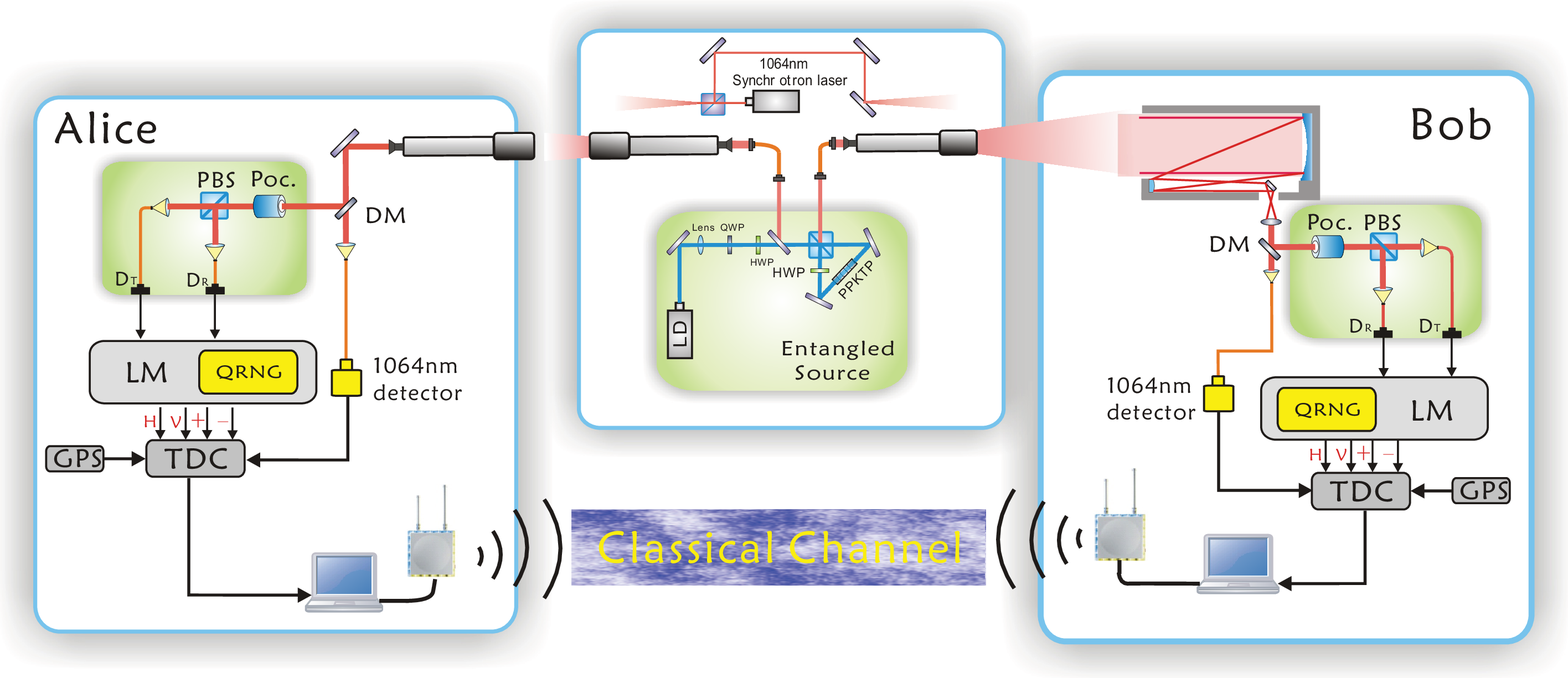}
\caption{A schematic diagram of the entanglement-based QKD setup. At the transmitter, a Sagnac PDC source generates entangled photon pairs. Two 80-mm-diameter telescopes are used to transmit photon pairs to two receiver sites, $7.8$ km away from the transmitter each. The detector modules, employed in both receivers, consist of a Pockels Cell (PoC), a polarization beam splitter cube (PBS) with two collimators placed in the transmitted and the reflected output port of PBS.}
\label{Fig:BiasedQKDExp:Setup}
\end{figure}

The two receiver sites, 15.3 km apart, are located $7.8$ km away from the transmitter, as shown in Fig.~\ref{Fig:BiasedQKDExp:Setup}. At Alice's receiver, settled in an observation platform of a hillside, we use a commercial 127-mm-diameter telescope to collect the entangled photons flying from the PDC source. At Bob's receiver, we use an off-axis parabolic telescope with a diameter of 400 mm. The collected photons are coupled into 105-$\mu$m-diameter multimode fibers with $0.22$ numerical apertures, and then detected by two single-photon counting modules (SPCMs) single-photon detectors. We separate the synchronization light (1064 nm) from the entangled photons (811 nm) via the dichroic mirror (DM) at both receivers. A $3$ nm narrow-band filter is used in front of each collimator to reduce the background noise. With the combined effect of the DM and the  filter, the background rate of each detector is less than $100$ counts per second. In addition, we utilize wireless bridge to exchange classical information between Alice and Bob.

In both receivers, Pockels Cells (PoCs), composed of potassium dideuterium phosphate (KD*P) crystals, are used to actively choosing polarization measurement bases. In order to make the PoC work as a switchable half-wave plate (HWP) for the basis choice ($X$ or $Z$ basis), we align the optical axis of PoC to $22.5^{\circ}$ in front of a polarization beam splitter (PBS).

\subsection{Basis bias setting}
On the receiver end, as shown in Fig.~\ref{Fig:BiasedQKDExp:Setup}, we use a PoC to actively choose the measurement basis. The tailored logic modules (LM) contains a quantum random number generater (QRNG), a complex programmable logic device (CPLD) chip and a single-chip microcomputer. The QRNG, with a maximum generation rate of $4$ MHz, controls a PoC to actively choose polarization measurement basis. The advantage of using active basis choice is that we can set an arbitrary basis bias flexibly. For the passive method, on the other hand, we need a specialized biased beam splitter, whose bias is normally fixed, for the efficient BB84 protocol.

\begin{figure}[hbt]\centering
  \includegraphics[width=10cm]{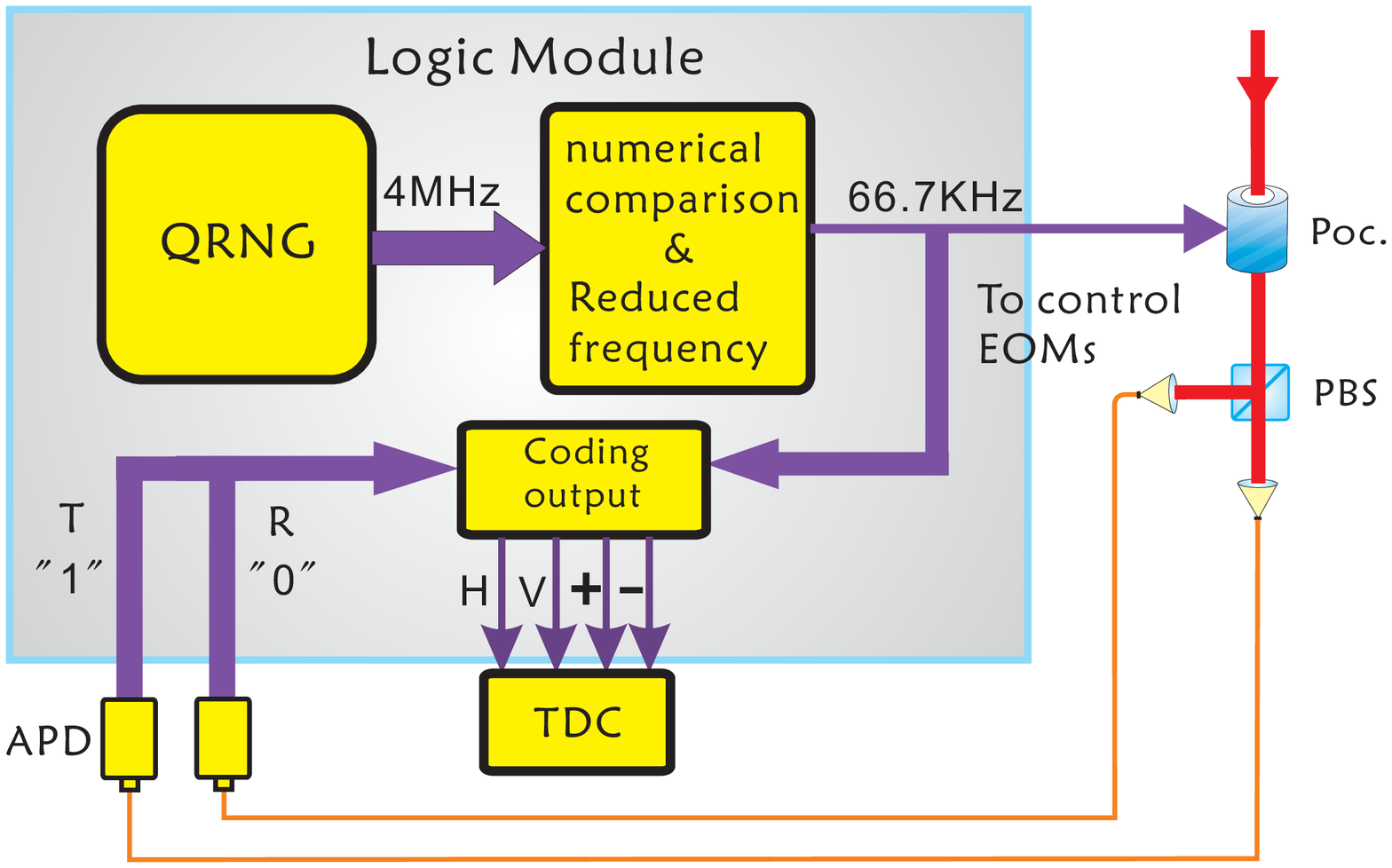}
\caption{A Logic Module is applied for controlling the Pockels Cells, switching polarization bases and outputting measurement results. Commercial QRNGs are applied to choosing measurement basis. The logical operation and the human-computer interface are realized via a CPLD (Complex programmable logic device) chip and a Single-Chip Microcomputer.} \label{Fig:BiasedQKDExp:LM}
\end{figure}

Figure \ref{Fig:BiasedQKDExp:LM} illustrates the workflow of LM. The first function of LM is choosing a basis bias. We load a 10-bit random number from the QRNG into buffer memory of our LM, denoted by $N_r$, and compare it to reference number, $N_0$. If $N_r > N_0$, LM outputs logic ``1", and else outputs logic ``0". Then the expected probability of outputting random number ``1" is
\begin{equation} \label{Biased:LM:Probability of q}
\begin{aligned}
Prob(a_{k}=1)=N_{0}\times2^{-10}.
\end{aligned}
\end{equation}
Now we can control the sifting factor $q$ through changing the number $N_{0}$. The controlling accuracy depends on the number of random bits to generate $N_{0}$.

The second function of LM is to distinguish the detected photons between four polarization states. The LM outputs a low-voltage modulation signal into a high-voltage pulser (HV1000) to switch the PoCs between half-wave voltage (about $+600V$) and biased voltage (about $+100V$). Part of the low-voltage modulation signal, which contains the basis information, and the output signal from APDs, which contains the bit information, are processed into one of the four states ${00, 01, 10, 11}$, representing the quantum states ${\left| H \right\rangle, \left| V \right\rangle, \left| + \right\rangle, \left| - \right\rangle}$, respectively. The final result is recoded using a time-to-digital converter (TDC), as shown in Fig.~\ref{Fig:BiasedQKDExp:LM}.

\subsection{Synchronization}
\begin{figure}[hbt]\centering
\includegraphics[width=12cm]{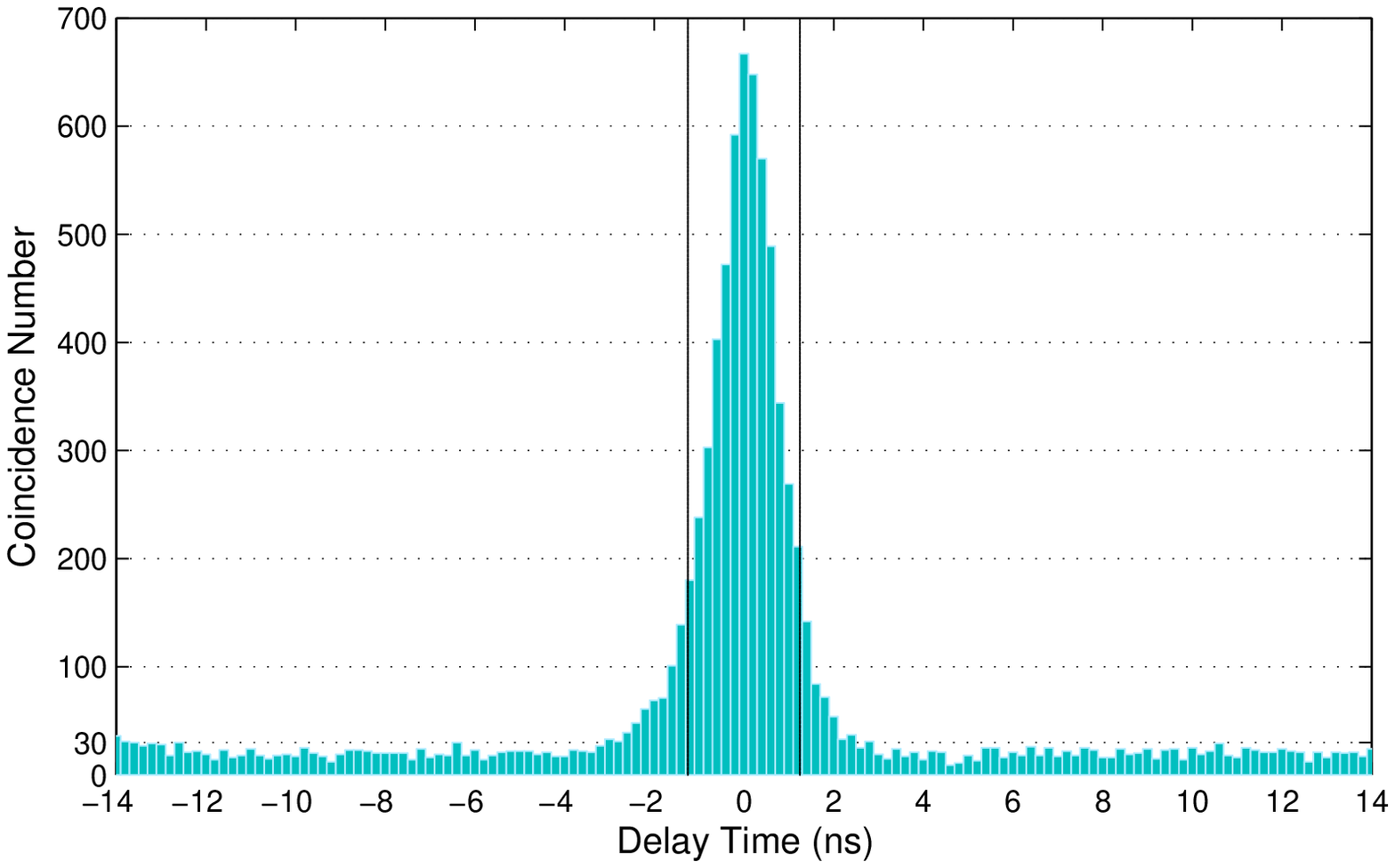}
\caption{The coincident detection events between Alice and Bob in a 25-min continuous sampling test. After balancing the number of coincident events and the source visibility, we set the gate width to be $2.5$ ns in our experiment.}
\label{Fig:BiasedQKDExp:syn}
\end{figure}

In order to achieve a high signal-to-noise ratio, we need a narrow coincidence time window to reduce the background count rate. We use a high-accuracy TDC to record the arrival time of the entangled photons at both receivers. For coarse synchronization, we employ the global position system (GPS). Then, we mainly follow the fine synchronization method used in \cite{PhysRevLett.94.150501}. Finally, we observe a coincidence peak with a FWHM around 1 ns. Coincidence time window of $2.5$ ns is used in the postprocessing of this experiment (see Fig.~\ref{Fig:BiasedQKDExp:syn}).

\section{Postprocessing and result} \label{Sec:Biased:Post}
The experiment lasted for nearly 3 hours at nighttime, with $10206$ seconds of effective QKD time. During the process, the average channel loss, taking into account of the transmitting and collection efficiency of the telescopes, the free-space channel loss and the fiber coupling efficiency, is around $50$ dB, including $29$ dB loss of Alice's optical link and $21$ dB of Bob's. Note that there is an extra 1$\sim$2 dB loss introduced by gate coincidence match and synchronization. By considering the balance between the signal-to-noise ratio and the raw key size, we set the coinciding time-gate to be $2.5$ ns. The results of the postprocessing are shown in Tab.~\ref{Tab:QKDExp:Result}.

\begin{table}[hbt]
\centering
\arraycolsep=9pt
\renewcommand{\arraystretch}{1.5}
\caption{List of measurement and post-processing results. After taking into account of error correction and privacy amplification, we evaluate the final secure key length. The sifted key length are $n_x$ and $n_z$ in the $X$ and $Z$ bases, respectively. The error correction efficiencies, $f_{ebx}$ and $f_{ebz}$, are evaluated by considering the low-density parity-check (LDPC) codes and progressive edge-growth (PEG) algorithm \cite{XiaoYuHu:LDPC:2005}. The failure probability $\varepsilon_{ph}$ in phase error rate estimation processing is defined in~\cite{Ma2011172}. The bit error rates in the $X$ and $Z$ bases are $e_{bx}$ and $e_{bz}$, respectively. Denote $q_{act}$ to be the experimental value of basis bias and $q_{opt}$ to be the optimal value given by theoretical calculation.} \label{Tab:QKDExp:Result}
\begin{tabular}{ccccccc}
  \hline
  Raw key & $n_x $ & $n_z $ & $f_{ebx}$ & $f_{ebz}$ & $\varepsilon_{ph}$&$\theta_{X}$\\
  \hline
  34644 & 1395 & 22300 & 1.1 & 1.12 & $6\times10^{-3}$ & 0.02\\
  \hline
  \hline
  $\theta_{Z}$ & $e_{bx}$ & $e_{bz}$ & $q_{act}$ & $q_{opt}$ & Final key & Improvement\\
  \hline
  0.019 & 0.069 & 0.065 & 0.8 & 0.79 & 4293  & 14.8\% \\
  \hline
\end{tabular}
\end{table}

In the calculation of the final secure key length, we employ the post-processing method proposed in~\cite{Ma2011172}, taking into account the finite-key size effect. As pointed in~\cite{Ma2011172}, the number of secure key bits cost in authentication, error verification, and the efficiency of privacy amplification is in the order of hundreds of bits. Thus, we neglect this part in our final key evaluation. Define $e_{bx}$ ($\delta_{px}$) and $e_{bz}$ ($\delta_{pz}$) to be the bit (phase) error rates in the $X$ and $Z$ bases, respectively. In the case of basis-independent source and long key limit, one can have \cite{KoashiPreskill_03},
\begin{equation} \label{Biased:postprocess:ebx}
\begin{aligned}
{\delta _{{\rm{px}}}} &= {e_{bz}}, \\
{\delta _{{\rm{pz}}}} &= {e_{bx}}.
\end{aligned}
\end{equation}
After performing the error correction and privacy amplification, the final secure key length is given by
\begin{equation} \label{Biased:postprocess:finalkey}
NR \ge {n_{sift}} - {k_{ec}} - {k_{pr}},
\end{equation}
where $NR$ is the final secure key length, $n_{sift}$ is the sifted key length, $k_{ec}$ is the secure-key cost of error correction, and $k_{pr}$ is the secure-key cost of privacy amplification. In the efficiency BB84 protocol, above parameters are given by,
\begin{equation} \label{Biased:postprocess:nsift}
\begin{aligned}
n_{sift} &= {n_x} + {n_z}, \\
k_{ec} &= {n_x}f({e_{bx}})H({e_{bx}}) + {n_z}f({e_{bz}})H({e_{bz}}), \\
k_{pr} &= {n_x}H({e_{bz}} + {\theta _z}) + {n_z}H({e_{bx}} + {\theta _x}), \\
\end{aligned}
\end{equation}
where $n_{x}$ and $n_{z}$ are the number of sifted bits produced in the $X$ and $Z$ bases, respectively, $f(x)$ is the error correction inefficiency, and $H(x) =  - x\log_2 x - (1 - x)\log_2 (1 - x)$ is the binary entropy function. When Alice and Bob estimate the phase error rates in $X$ and $Z$ bases, $\theta_{x}$ and $\theta_{z}$ are the corresponding deviations due to statistical fluctuations. We evaluate these deviations as follows. Take $Z$ basis as an example. Define the probability of ${e_{pz}} > {e_{bx}} + {\theta _x}$ to be $P_{\theta_{x}}$,
\begin{equation} \label{Biased:postprocess:ptheta}
{P_{{\theta _x}}} < \frac{{\sqrt {{n_x} + {n_z}} }}{{\sqrt {{n_x}{n_z}{e_{bx}}(1 - {e_{bx}})} }}{2^{ - ({n_x} + {n_z}){\xi _x}({\theta _x})}}
\end{equation}
where ${\xi _x}({\theta _x}) \equiv H({e_{bx}} + {\theta _x} - {q_x}{\theta _x}) - {q_x}H({e_{bx}}) - (1 - {q_x})H({e_{bx}} + {\theta _x})$ with $q_{x}=n_{x}/(n_{x}+n_{z})$. Similar results hold for $Z$-basis case, $P_{\theta_{z}}$. Then the total failure probability of phase error rate estimation, $\varepsilon_{ph}$, is given by
\begin{equation} \label{Biased:postprocess:ptheta}
{\varepsilon _{ph}} \le {P_{{\theta _x}}} + {P_{{\theta _z}}}
\end{equation}
In the post processing, we obtain the error rates $e_{bz}$, $e_{bx}$ and the raw key length $N$ from the experiment, and fixed the failure probability $\varepsilon_{ph}$ to be a small number. Here the failure probability we chose for $\varepsilon_{ph}$ is 0.3\% in both two bases X and Z.

The result is listed in Table \ref{Tab:QKDExp:Result}. We use a bias of $20:80$ ($X:Z$), which is close to the theoretical optimal value of $21:79$. In the asymptotic case, where the finite-key effect is neglected, the final key size should achieve an $36\%$ increase over the unbiased case \cite{improvement}. After considering the finite-key effect, we obtain a final key rate of $0.124$ bit/raw key.

For comparison, we fix the raw key size to simulate the unbiased case by dividing equally the sifted keys into two bases and keep all the other parameters unchanged and then we obtain the final key rate of $0.108$ bit/raw key. Hence, we increase the final key rate by $14.8\%$ through implementing the efficient BB84 protocol. Comparing to the infinite-key improvement of $36\%$, one can see that the finite-key analysis does affect the QKD data postprocessing. The reason for this gap is mainly due to the relatively small final key size ($4293$ bits in our experiment). Therefore, we emphasize the importance of finite-key-size analysis, especially in the case of a small final key size (such as $< 10^{5}$). This result is compatible to the theoretical simulations \cite{Ma2011172}.

In fact, if we had obtained a longer key in our experiment (say, by extending the experimental time or increasing the telescope aperture to reduce the channel loss), such as $1$ Mbits raw key, we could have achieved key-rate improvement of $71\%$ (by taking a theoretical optimal $4:96$ bias) over the unbiased case with the finite-key effect taken into account. Therefore, we conclude that the efficient BB84 protocol is an effective way to increase the final secrete key rate.

The final secure key rate is low in this experiment (about $0.42$ bit/s). There are mainly two reasons. First, the key rate is limited by the brightness of the entangled photon source. Recently, a lot of efforts have been devoted to improve the brightness of entangled photon sources~\cite{Steinlechner:highBrigthnessES:2012}. Second, the quantum efficiency of free-space channels are limited by the photon collection efficiency, which are greatly affected by beam divergence and atmosphere turbulence. To improve the photon collection efficiency, one can employ a wide-aperture telescope and equip a high-frequency and high-accuracy acquiring, pointing and tracking system. Some of the improvement techniques have been experimental demonstrated recently \cite{yangbin:decoystate:2013}.

\section{Conclusion and discussions} \label{Sec:Biased:Conclusion}
We have experimentally demonstrated the free-space entanglement-based QKD with the efficient BB84 protocol. The final key rate in our biased basis setting is increased by $14.8\%$ comparing to the standard BB84 protocol, after considering the finite-key effect. The experimental result suggests that the efficient BB84 protocol is indeed an easy and effective way to increase the key rate for QKD. Moreover, the developed entanglement source, active basis choice and synchronizing technology in this experiment are necessary components for future satellite-based quantum communication.

Note that a couple of feasibility tests for ground-satellite QKD have been done recently \cite{Nauerth:airtoground:2012,yangbin:decoystate:2013}, both of which employ the prepare-and-measure QKD protocol. Lately, another type of scheme, the entanglement-based QKD where an entangled source is put in a satellite and signals are transmitted to two ground stations via two quantum channels, attracts lots of research interests \cite{Xin:ScienceNews:2011,Pan:ScienceNews:2012}.
Considering that some properties (such as the channel lengths of 7.8 km and the total channel loss of 50 dB) of our two free-space optical links are comparable to the down-links between a low Earth orbit satellite and an optical ground station \cite{Bonato:FeasibilitySatQKD:NJP2009,Yin:SinglephotonSat:OE2013}, our experiment can be regarded as a partial simulation of the entanglement-based ground-satellite QKD.

In our setup, the basis bias can be randomly modified in real-time. It is interesting to study whether one can take advantage of this extra degree of freedom of basis choice. Also, the basis is actively chosen in our setup, which allows us to apply four-state polarization analyzer \cite{USpa_Mismatch_05} to foil some quantum attacks, such as the time-shift attack \cite{Qi:TimeShift:2007,Zhao:TimeshiftExp:2008} due to the efficiency mismatch of detectors \cite{Makarovet:PhysRevA:efficiencymismatch}. This can be done by adding one more PoC with $45^{\circ}$ aligned to the optical axis in Fig.~\ref{Fig:BiasedQKDExp:Setup}. Since the single photon detectors operate continuously, one of the potential advantages is that the attacking strategy presented in ~\cite{PhysRevLett.107.110501}, called calibration loophole, is not effective for our QKD system. Furthermore, the technique of actively choosing measurement bases developed in this work can also be applied to  other scientific experiments, such as the fundamental tests of quantum mechanics --- nonlocal correlation \cite{Yin:Spooky:2013}.

In a passive basis choice setting, such as the one used in~\cite{Erven:Biased:2009}, there is a potential drawback compared to our active setting as we used here. In the passive setting, four detectors (two for each basis) always open regardless of basis choices. Hence, when a large bias is used, the effect of dark counts on the basis with a lower rate may increase. For example, suppose we use a bias of 10:90 for the $X$/$Z$-basis, then the number of errors introduced by dark counts in $X$-basis is 10 times higher than the one in the active setting.

\section*{Acknowledgments}
We acknowledge Wei-Yue Liu, Yang Liu and Yan-Ling Tang for useful discussions in the error correction part. We thank Qiang Zhang, Yu-Ao Chen and Bo Zhao for their valuable suggestions. This work has been supported by the National Fundamental Research Program (under grant No.~2011CB921300 and No.~2013CB336800 ), the Chinese Academy of Sciences and the National Natural Science Foundation of China, the National Basic Research Program of China Grants No.~61073174, No.~61033001,No.~61061130540, No.~2011CBA00300 and No.~2011CBA00301, the 1000 Youth Fellowship program and the Chinese Academy of Sciences.


\begin{thebibliography}{10}
\bibitem{improvement}
This can be simply calculated by comparing the basis-sift factor of $0.2^2+0.8^2=0.68$ to 1/2.
\end{thebibliography}
\end{document}